\documentclass{ws-mpla}

\begin{document}

\markboth{C. H. Hyun et al.}
{The old and the new of parity-violating two-pion-exchange $NN$ potential}

\catchline{}{}{}{}{}

\title{
THE OLD AND THE NEW OF PARITY-VIOLATING TWO-PION-EXCHANGE $NN$ POTENTIAL
}

\author{C. H. HYUN}

\address{Department of Physics Education,
Daegu University, Gyeongsan 712-714, Korea \\
hch@color.skku.ac.kr}

\author{B. DESPLANQUES}

\address{
LPSC, Universit\'e Joseph Fourier Grenoble 1, CNRS/IN2P3, INPG, \\
  F-38026 Grenoble Cedex, France
}

\author{S. ANDO}

\address{
Theoretical Physics Group,
School of Physics and Astronomy, \\
The University of Manchester, Manchester, M13 9PL, UK
}

\author{C.-P. LIU}

\address{
Department of Physics, University of Wisconsin-Madison, \\
1150 University Avenue, Madison, WI 53706-1390, USA
}

\maketitle

\pub{Received (Day Month Year)}{Revised (Day Month Year)}

\begin{abstract}
We consider the parity-violating two-pion-exchange potential 
obtained from the covariant formalism in the past 
and the state-of-the-art effective field theory approach.
We discuss the behavior of the potential in coordinate space 
and its application to the parity-violating
asymmetry in $\vec{n} p \rightarrow d \gamma$ at threshold.

\keywords{Nucleon-nucleon interaction; Applications of electroweak
models to specific processes}
\end{abstract}

\ccode{PACS Nos.: 13.75 Cs, 12.15 Ji}

\section{Introduction}	

%
Effective field theory (EFT) approach, which shed new light on the
methodology of nuclear physics, was used to reformulate the parity-violating
(PV) nucleon-nucleon ($NN$) interaction recently,\cite{zhu05} and the 
potential has been applied to a few two-nucleon problems.\cite{had07,cpl07}
The PV two-pion-exchange (TPE) contribution to physical observables
turned out to be non-negligible, but was not so significant as the TPE
in the strong $NN$ interaction. On the other hand, a PV $NN$ contact term
parametrized by a low-energy constant (LEC), which is presumed to 
subsume the heavy degrees of freedom and
higher-order corrections, is comparable to the PV TPE contribution.
More importantly, the LEC term (the contact term) is highly singular 
at short distances and depends critically on the regularization methods 
and renormalization schemes.
With the lack of experimental data to determine the LEC value, it is not
feasible to sharpen the theoretical prediction furthermore.

As a partial resolution to the uncertainty associated with the LEC term, 
we consider the covariant formulation of the PV TPE potential, which was 
already carried out in 1970's.\cite{bd72,pirner73,bd74}
Compared to the TPE and LEC terms in EFT, the covariant formalism gives 
a potential which is finite and has no unknown constant.
Moreover, it is less singular at short distances than that from EFT,
and contains more higher-order $1/M$ contributions.
In this work we focus on comparing the TPE potential from the covariant
formalism with that from EFT. By calculating a PV asymmetry in
$\vec{n} p \rightarrow d \gamma$ with these potentials, we estimate
the correction from the higher-order terms embedded in the covariant
potential. The analysis will provide an indirect estimation of the
correction due to the LEC term in EFT.

\section{Formalism}

The PV TPE potential contains various spin and isospin components.
Keeping the spin and isospin operators only relevant for the asymmetry 
$A_\gamma$ in $\vec{n} p \rightarrow d \gamma$, we get the following
expression :
\begin{eqnarray}
V^{\vec{n}p}_{\mbox{\tiny TPE}}(\vec{r}) = 
i (\vec{\tau}_1 \times \vec{\tau}_2)^z \, (\vec{\sigma}_1 + \vec{\sigma}_2)
\cdot \left[ \vec{p},\, v_{56}(r) \right] +
(\tau^z_1 - \tau^z_2) \, (\vec{\sigma}_1 + \vec{\sigma}_2) \cdot
\left\{ \vec{p},\, v_{75}(r) \right\}.
\end{eqnarray}
The radial part of the potential in the covariant approach can be written
in the form 
\begin{eqnarray}
v^{\mbox{\tiny cov}}_{ij}(r) = \frac{1}{4\pi^2} \int^\infty_{4 m^2_\pi} dt'
g_{ij}(t') \frac{e^{-r \sqrt{t'}}}{r \sqrt{t'}},
\end{eqnarray}
where $m_\pi$ is the pion mass and $g_{ij}(t')$'s are the
spectral functions for given indices $i$ and $j$.
%
%
We have $g_{56}$ and $g_{75}$ as
\begin{eqnarray}
g_{56}(t')&=& \frac{g^3_{\pi NN} h^1_\pi}{32 \sqrt{2} \pi} \left[
-\frac{1}{2M}\;\frac{H\;x}{x^2+4M^2q_{\pi}^2}
-\frac{x}{M^2 \; m^2_{\pi}}\;
 {\rm arctg} \Big (\frac{m^2_{\pi}}{2Mq_{\pi}}\Big ) \right.
\nonumber \\ 
& & \left. + \int_{k_{-}^2}^{{k_{+}^2}} 
\frac{dk^2}{\sqrt{k^2t'\!-\!(m^2_{\pi}\!+\!k^2)^2}}\;
\frac {1}{2E\,(E\!+\!M)} 
\Bigg (\frac {x}{M^2}\!+\!\frac {k^2}{M\,(E\!+\!M)} \Bigg )
\right],
\nonumber \\
g_{75}(t')&=& \frac{g^3_{\pi NN} h^1_\pi}{32 \sqrt{2} \pi} \left[
\frac{4x^2}{M^2 \, m^2_{\pi}\,t'}\;
 {\rm arctg} \Big (\frac{m^2_{\pi}}{2Mq_{\pi}}\Big )+ \frac{2G}{M t'}
\right.
\nonumber \\ 
&-& \left. \int_{k_{-}^2}^{{k_{+}^2}} 
\frac{dk^2}{\sqrt{k^2t'\!-\!(m^2_{\pi}\!+\!k^2)^2}}
\frac{2}{E(E\!+\!M)}\Bigg (
\frac{x^2 + 2E M x}{M^2 t'}
\!-\! \frac{k^2\, t' + k^4}{M(E\!+\!M)t'}  \Bigg )\right],
\label{eq:gcad}
\end{eqnarray}
where $M$ is the nucleon mass and $h^1_\pi$ the weak $\pi NN$ coupling
constant. The definitions of the functions $q_\pi$,
$\chi^2$, $x$, $H$, $G$, $k^2_\pm$ and $E$, which are non-linear functions
of $t'$ and $M$, can be found in Ref.~7. 
Due to the non-linearity of these functions, 
when expanded in terms of $1/M$,  
$g_{56}$ and $g_{75}$ have infinite terms in the power of $1/M$.
In order to understand the relation between the covariant and
EFT approaches, we take the large nucleon mass (LM) limit for
the spectral functions. Keeping only the leading terms of $1/M$ 
in each function, we have
\begin{eqnarray}
\lim_{M\rightarrow \infty} g_{56}(t') = 
- \frac{g^3_{\pi NN} h^1_\pi}{32\sqrt{2} \pi} \frac{x}{q_\pi M^3},\,\,\,
\lim_{M\rightarrow \infty} g_{75}(t') =
\frac{g^3_{\pi NN} h^1_\pi}{32\sqrt{2} \pi} 
\frac{\pi}{4 M^4}\left(q^2_\pi+3x\right).
\end{eqnarray}
For $g_{56}$ the leading order is $1/M^3$; for $g_{75}$, it is $1/M^4$.
Consequently, $v_{56}$ becomes the leading term in LM, 
and $v_{75}$ is higher-order in $1/M$.
We have explicit form for $v_{56}$ in momentum space as
\begin{eqnarray}
v^{\mbox{\tiny LM}}_{56}(q) = -\sqrt{2} \pi \frac{g^3_{\pi NN} h^1_\pi}{
(4 \pi g_A M)^3} \left[g^3_A L(q) - g^3_A \left(3 - \frac{4m^2_\pi}{4m^2_\pi
+q^2}\right) L(q) \right],
\label{eq:v56lm}
\end{eqnarray}
where $g_A$ is the axial coupling constant and the definition of
$L(q)$ can be found in Ref.~2.
The component
$v^{\mbox{\tiny LM}}_{56}$ given by Eq.~(\ref{eq:v56lm}) is almost the 
same as the corresponding TPE term in EFT, except for the factor 
$g^3_A$ of the first term in the square bracket, which is $g_A$ in 
EFT.\cite{had07} 
The discrepancy is explained in Ref.~7, and it is concluded that
the leading $1/M$ term extracted from the covariant form is 
in principle equivalent to the TPE term obtained from EFT.
In the next section, we present numerical results for the potentials
and their application.

\section{Numerical Results}


\begin{figure}[ph]
\centerline{\psfig{file=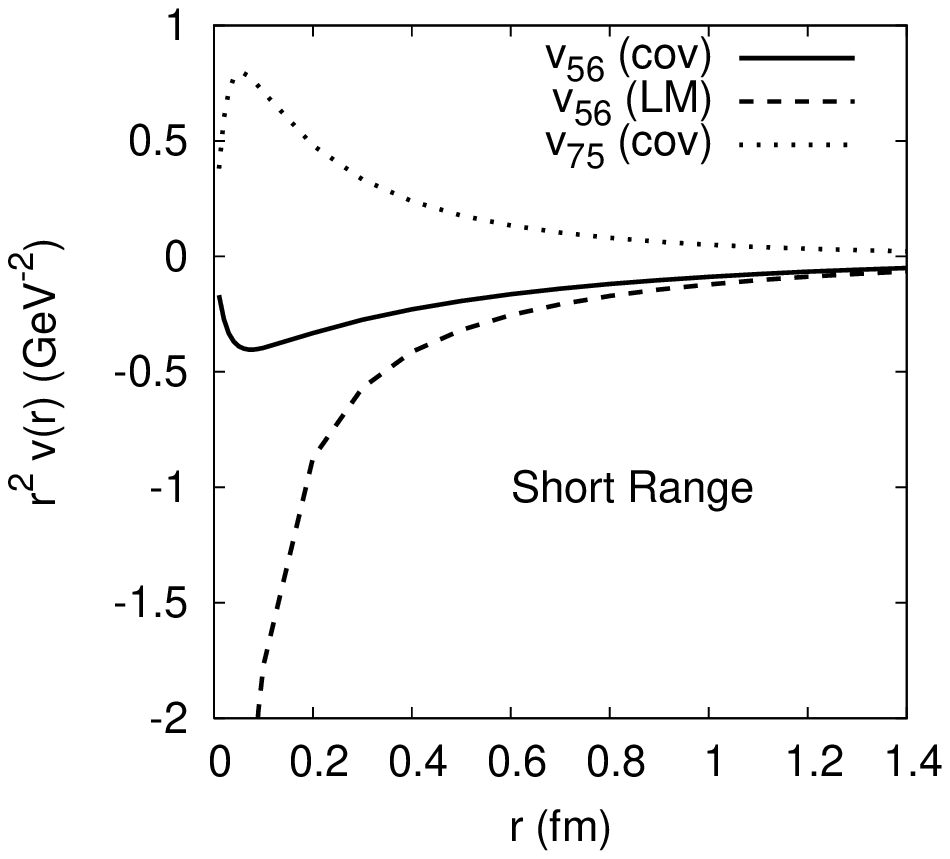,width=6cm}
\psfig{file=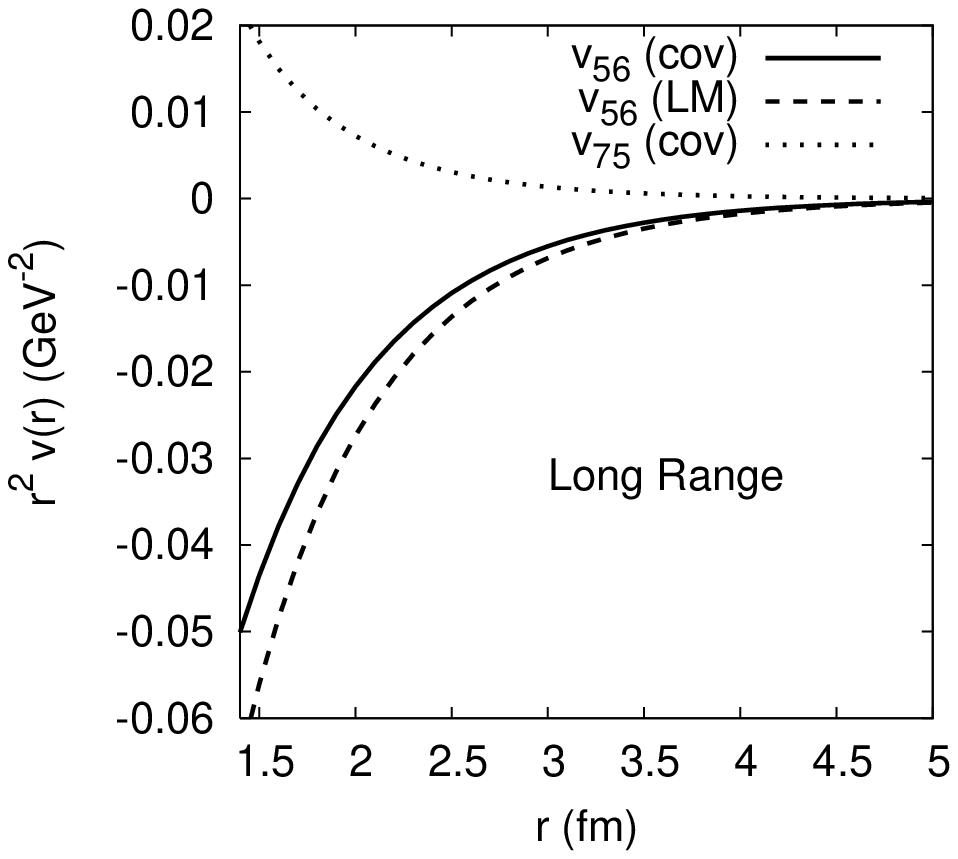,width=6cm}}
\caption{TPE potentials in the short (left) and the long (right) range region.
\protect\label{fig}}
\end{figure}

Fig.~\ref{fig} shows the $v_{56}$ and $v_{75}$ components of the
PV TPE potential in coordinate space multiplied by $r^2$.
In the short-range region (left panel), the covariant terms
$v_{56}$ and $v_{75}$ both show non-singular behavior around $r=0$, 
while the LM term diverges as $r \rightarrow 0$. 
We expect that when a properly renormalized LEC term is added to 
$v^{\mbox{\tiny LM}}_{56}$, it will make the net behavior less singular
at small $r$. On the other hand, the covariant and LM forms show convergent 
behavior in the long-range region (right panel).


The PV asymmetry $A_\gamma$ in $\vec{n} p \rightarrow d\gamma$ is defined
as the photon asymmetry with respect to the neutron polarization. 
Treating $h^1_\pi$ perturbatively and keeping only the leading-order
contribution, $A_\gamma$ is proportional to $h^1_\pi$, 
i.e. $A_\gamma = a_\gamma h^1_\pi$.
The factor $a_\gamma$ depends on both strong and weak dynamics.
In order to reduce the uncertainty from the strong interaction, we employ 
the Argonne v18 potential to obtain the parity-conserving part of the 
wave function.
Numerical results for $a_\gamma$ are summarized in Table~\ref{table}.
The covariant contributions from $v_{56}$ and $v_{75}$ are summed to cancel 
strongly, $0.0093-0.0040 = 0.0053$, which is much smaller than the 
one-pion-exchange (OPE) one, $-0.1120$. Comparing the contribution of
$v^{\mbox{\tiny cov}}_{56}$, which contains infinite orders of $1/M$, with 
$v^{\mbox{\tiny LM}}_{56}$, which is proportional to $1/M^3$, 
the higher-order $1/M$ corrections are non-negligible, but less significant 
than the leading $1/M$ contribution.

\begin{table}[tbp]
\tbl{Numerical results for the asymmetry. $v_{\mbox{\tiny OPE}}$ represents
the PV one-pion-exchange (OPE) potential.}
{\begin{tabular}{@{}ccccc@{}} \toprule
\phantom{0} Potential \phantom{0} & 
\phantom{0} $v_{\mbox{\tiny OPE}}$ \phantom{0} &
\phantom{0} $v^{\mbox{\tiny cov}}_{56}$ \phantom{0} & 
\phantom{0} $v^{\mbox{\tiny cov}}_{75}$ \phantom{0} & 
\phantom{0} $v^{\mbox{\tiny LM}}_{56}$ \phantom{00} \\
\colrule
$a_\gamma$  & $-0.1120$ & 0.0093 & $-0.0040$ & 0.0141 \\
\colrule
\end{tabular}}
\label{table}
\end{table}

\vspace*{10pt}
In summary, the behavior of the potential at short distances
and the PV asymmetry in $\vec{n} p \rightarrow d\gamma$ show that 
the uncertainty due to the 
corrections from the degrees of freedom integrated out of the model
is still significant, and it requires further investigation. 
However, the TPE and higher-order contributions to $A_\gamma$, 
in either covariant or EFT form, gives a correction about 10~\% at most. 
Recent calculations of the $\Delta(1232)$ resonance
contribution to the PV potential in EFT show that the present result of 
$A_\gamma$ is not much affected by the $\Delta$ contribution.\cite{kaiser07}
To the extent investigated so far, $A_\gamma$ is dominated
by the PV OPE potential with an uncertainty of about 10\% at most. 
The precise measurement of $A_\gamma$ will thus provide a unique 
opportunity to determine the value of $h^1_\pi$. 

\section*{Acknowledgments}

Work of CHH was supported by the Korea Research Foundation Grant 
funded by the Korean Government (MOEHRD, Basic Research Promotion
Fund) (KRF-2007-313-C00178). SA is supported by STFC grant number
PP/F000448/1.

\end{document}